\documentclass[12pt]{iopart}
\usepackage{graphicx}

\providecommand{\draftonly}[1]{}

\newcommand{\beq}{\begin{eqnarray}}
\newcommand{\eeq}{\end{eqnarray}}
\newcommand{\nel}{\nonumber\\}
\newcommand{\aeq}{\\}
\newcommand{\brar}[1]{\left(#1\right)} 
\newcommand{\brac}[1]{\left\{#1\right\}} 
\newcommand{\braa}[1]{\left\langle#1\right\rangle} 
\newcommand{\intzi}{\int_0^\infty}
\newcommand{\conj}[1]{{#1}^\ast}
\newcommand{\adj}[1]{{#1}^\dag}
\newcommand{\modulus}[1]{\left|#1\right|}
\newcommand{\modsquared}[1]{{\modulus{#1}^2}}
\newcommand{\dd}[1]{\frac{d}{d#1}}
\newcommand{\ddt}{\dd{t}}
\newcommand{\com}[2]{\left[#1,#2\right]}
\newcommand{\half}{\frac{1}{2}}

\newcommand{\bra}[1]{\left\langle#1\right|}
\newcommand{\ket}[1]{\left|#1\right\rangle}
\newcommand{\ex}[1]{\braa{#1}}
\newcommand{\op}[2]{\ket{#1}\bra{#2}}
\newcommand{\proj}[1]{\op{#1}{#1}}
\newcommand{\dampingnobr}[3]{2#1#2#3-#3#1#2-#2#3#1} 
\newcommand{\damping}[3]{\brar{\dampingnobr{#1}{#2}{#3}}}
\newcommand{\lio}{\mathcal{L}} 
\newcommand{\ad}{\adj{a}}
\newcommand{\sigm}{\sigma^-}
\newcommand{\sigp}{\sigma^+}
\newcommand{\sigx}{\sigma_x}
\newcommand{\sigy}{\sigma_y}
\newcommand{\sigz}{\sigma_z}
\newcommand{\sigmp}{\sigm\sigp}
\newcommand{\sigpm}{\sigp\sigm}

\newcommand{\commentout}[1]{\draftonly{\textit{#1}}}

\newcommand{\percent}{\%}

\begin{document}
\title{Coupling of effective one-dimensional two-level atoms to squeezed light}
\author{Stephen Clark and Scott Parkins}
\address{Department of Physics, University of Auckland, Private Bag 92019, Auckland, New Zealand.}
\date{\today}

\begin{abstract}
A cavity QED system is analyzed which duplicates the dynamics of a two-level atom in free space interacting exclusively with broadband squeezed light. We consider atoms in a three or four-level $\Lambda$-configuration coupled to a high-finesse optical cavity which is driven by a squeezed light field. Raman transitions are induced between a pair of stable atomic ground states via the squeezed cavity mode and coherent driving fields. An analysis of the reduced master equation for the atomic ground states shows that a three-level atomic system has insufficient parameter flexibility to act as an effective two-level atom interacting exclusively with a squeezed reservoir. However, the inclusion of a fourth atomic level, coupled dispersively to one of the two ground states by an auxiliary laser field, introduces an extra degree of freedom and enables the desired interaction to be realised. As a means of detecting the reduced quadrature decay rate of the effective two-level system, we examine the transmission spectrum of a weak coherent probe field incident upon the cavity.
\end{abstract}
\submitto{\JOB}
\ead{\mailto{scla016@phy.auckland.ac.nz}, \mailto{s.parkins@auckland.ac.nz}}

\section{Introduction}
The interaction of squeezed light and atoms is a problem of fundamental interest in quantum optics~\cite{Par93,Dal99}. Of particular interest to us here is the prediction by Gardiner in 1986~\cite{Gar86} that an atom interacting exclusively with squeezed light would display an inhibited phase decay rate for one of the atomic polarization quadratures. This inhibited phase decay results in a feature with sub-natural linewidth in the fluorescent spectrum that is a direct indication of the noise reduction in one quadrature of the squeezed light field.

Unfortunately, to date, there has only been limited experimental investigation of the alteration of the fundamental radiative processes of atoms when interacting with squeezed light~\cite{Geo95,Tur98/2}. This is in part due to the difficulty of coupling the atom exclusively to squeezed modes of the incident light field. Here we analyze a scheme to create an effective two-level atom that can be considered to be \emph{one-dimensional}~\cite{Tur95/2}, i.e., it couples predominantly to a single field mode of a high-finesse optical cavity. Under appropriate conditions the damped cavity provides the dominant input and output channel to the atom. For the atom to interact primarily with squeezed modes then requires that only the external field modes which couple strongly to the cavity mode must be squeezed.

Experiments in cavity QED involve either beams of atoms traversing the cavity, or single atoms trapped within it. The major difficulty associated with the use of an atomic beam is the resulting uncertainty in the number of atoms interacting with the cavity mode. This was a major limiting factor in the squeezed excitation experiment of~\cite{Tur98/2}. Fortunately, there have been dramatic recent advances in the field of single-atom trapping in optical cavities~\cite{Ye99,Hoo00,Pin00,McK02P}; these advances provide motivation for this paper.

\begin{figure}
  \begin{center}
  \includegraphics{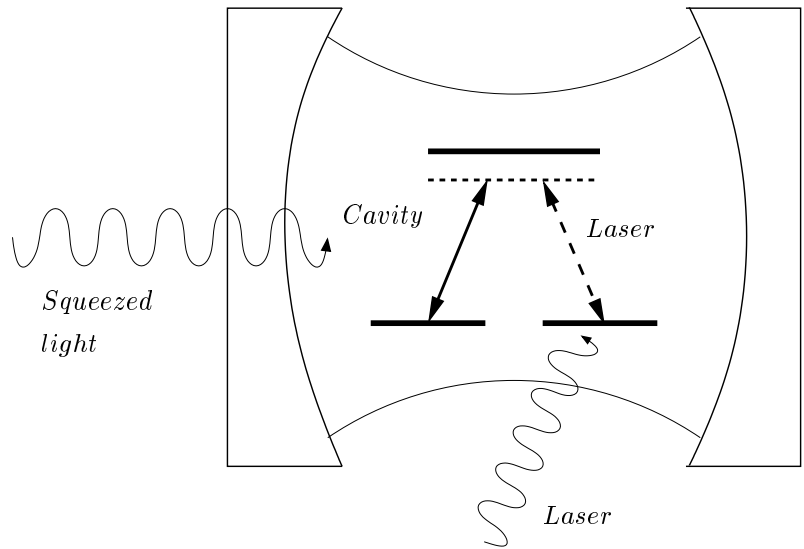}
  \end{center}
  \caption{Schematic of a 3-level $\Lambda$-atom in a cavity driven by squeezed light. The Raman transition is driven by a detuned coupling laser and the cavity mode.}
  \label{fig:cav}
\end{figure}

A single 3-level atom in the $\Lambda$-configuration, trapped in a high-finesse optical cavity, with a Raman transition between the ground states that is driven by a coupling laser and by the cavity mode is recognized as a possible implementation of a qubit (see, e.g.,~\cite{Cir97} for a proposed use of such systems in a quantum network). This system is shown schematically in figure~\ref{fig:cav}.
The aim of this paper is to establish parameter regimes under which a three-level atom in a squeezed optical cavity undergoes the same dynamics as a two-level atom in free space interacting exclusively with a squeezed vacuum.

The 3-level $\Lambda$-atom in a cavity may be analyzed by adiabatically eliminating first the excited state of the atom and then the cavity mode to obtain a reduced master equation governing just the evolution of the atomic ground states. It is in this context that we may refer to it as an effective two-level atom. The adiabatic elimination of the excited state of the atom may be easily performed by assuming a large detuning of the light fields driving the Raman transition. After this step we obtain an effective Jaynes-Cummings interaction between the ground states of the atom and the cavity mode. The adiabatic elimination of the cavity mode may then be undertaken under the assumption that the cavity decay rate dominates the characteristic rate associated with this interaction.

Two major differences arise between our 3-level $\Lambda$-system and an actual two-level atom. The elimination of the atomic excited state creates level shifts in the two atomic ground states caused both by the driving laser field and by intra-cavity photons. These two level shifts are not present in the two-level atom in free space and must be made identical to keep the atomic quadratures at a constant phase with respect to the external squeezing. In addition, the level shift in the atomic ground state that couples to the cavity mode causes a detrimental phase decay due to a coherent scattering process between the effective two-level atomic system and the intra-cavity photons.

Both of these differences must be made negligible to create an equivalence in the dynamics between the 3-level $\Lambda$-atom in a cavity and a two-level atom in free space (in the presence of squeezed light). We shall show that there is insufficient flexibility in the coupling and driving parameters to minimize the undesirable phase decay and simultaneously equalize the level shifts of the atomic ground states. Thus the three-level system driven by squeezed light in a cavity is not suitable for simulating the dynamics of a two-level atom interacting exclusively with squeezed light.

This problem may be overcome by making a small extension to employ a fourth atomic level and an additional driving laser to virtually excite it. In this new four-level atomic system we may minimize the effect of the coherent scattering process and simultaneously balance the level shifts in the two atomic ground states. Therefore, in the appropriate parameter regime, we show that this four-level system can be used to reproduce the dynamics of the simpler two-level atom, including the linewidth narrowing associated with reduced quadrature fluctuations.

It is also pertinent to note that the results in this paper may be straightforwardly extended to systems containing multiple atoms. It has been shown that the interaction of squeezed light with multiple two-level atoms can lead to entangled pure atomic states~\cite{Eke89,Pal89,Aga90} which have possible applications in quantum teleportation and quantum computation; this is also a reason for our interest in realising an effective two-level system with a pair of stable (long-lived) atomic ground states. Note that squeezed light beams have in fact already been used to induce weak spin-squeezing of large atomic ensembles~\cite{Kuz97,Hal99}.

The structure of this paper is as follows. In section~\ref{sec:a2} we briefly review the theoretical prediction of inhibited quadrature decay for a two-level atom interacting exclusively with a broadband squeezed vacuum. In section~\ref{sec:a3} we consider the dynamics of a three-level atom in a cavity and show that there is insufficient freedom in the choice of coupling and driving parameters to faithfully simulate an equivalent two-level atom interacting with a broadband squeezed light field. To provide more flexibility, we introduce in section~\ref{sec:a4} a slightly more complex four-level atomic scheme. A parameter regime is then identified in which this configuration has the required dynamics. For this system we include the effects of atomic spontaneous emission in our analysis and also demonstrate a technique, involving a pair of probe fields, to identify a particular atomic quadrature decay rate in the probe transmission spectrum.

\section{Two-level atom and broadband squeezed light}
\label{sec:a2}
The master equation for the density operator~$\rho$ describing a two-level atom with linewidth~$\gamma$, interacting resonantly (and exclusively) with a broadband squeezed vacuum characterized by the standard squeezing parameters~$N$ and~$M$ ($\modulus{M}=\sqrt{N(N+1)}$ for ideal squeezing), is~\cite{Gar85,Par93}
\beq
  \label{eq:a2:me}
  \dot{\rho}&=&
  \frac{\gamma}{2}\brar{N+1}
  \damping{\sigm}{\rho}{\sigp}
\nel
  &&+\frac{\gamma}{2}N
  \damping{\sigp}{\rho}{\sigm}
\nel
  &&-\gamma M\sigp\rho\sigp-\gamma\conj{M}\sigm\rho\sigm,
\eeq
where $\sigp$ ($\sigm$) is the atomic raising (lowering) operator. Defining $\sigx=\sigp+\sigm$, $\sigy=-i\brar{\sigp-\sigm}$ and $\sigz=\sigp\sigm-\sigm\sigp$, the equivalent Bloch equations are (taking~$M$ to be real) \commentout{(W107H)}
\beq
  \label{eq:a2:bex}
  \ddt\ex{\sigx}&=&-\gamma_x\ex{\sigx},
\aeq
  \label{eq:a2:bey}
  \ddt\ex{\sigy}&=&-\gamma_y\ex{\sigy},
\aeq
  \label{eq:a2:bez}
  \ddt\ex{\sigz}&=&-\gamma-\gamma_z\ex{\sigz},
\eeq
with decay rates
\beq
  \gamma_x&=&\frac{\gamma}{2}(2N+1+2M),
\aeq
  \gamma_y&=&\frac{\gamma}{2}(2N+1-2M),
\aeq
  \gamma_z&=&\gamma(2N+1).
\eeq
The decay rate of the $\ex{\sigy}$ quadrature may be strongly inhibited when squeezing parameters are chosen so that~$N$ is large and $M=\sqrt{N(N+1)}\approx N+\half$. The Bloch equations that we obtain for the cavity QED systems considered in sections~\ref{sec:a3} and~\ref{sec:a4} will be compared to equations~(\ref{eq:a2:bex}--\ref{eq:a2:bez}) to see if they have the same form.

\section{Three-level $\Lambda$-system in a cavity}
\label{sec:a3}
\begin{figure}
  \begin{center}
  \includegraphics{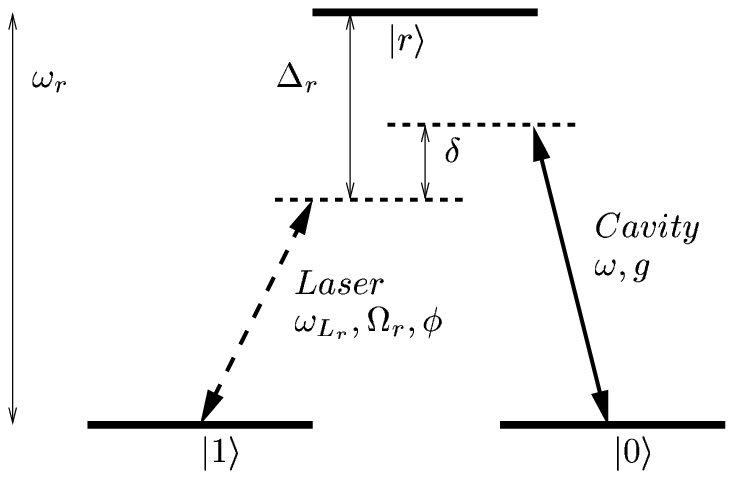}
  \end{center}
  \caption{Level scheme for a 3-level $\Lambda$-atom in a cavity. The Raman transition is driven by a coupling laser and the cavity mode. See the text for a description of the variables.}
  \label{fig:a3}
\end{figure}

In this section we examine the properties of a three-level $\Lambda$-atom trapped in a high-finesse optical cavity of (field) decay rate $\kappa$ and frequency $\omega$ (see figure~\ref{fig:a3}).
The annihilation (creation) operator for the quantized cavity mode is denoted $a$ ($\ad$). The cavity is driven by squeezed light characterized by the parameters $N$ (real) and $M$ (complex). The atom is assumed to have two stable ground states, $\ket{0}$ and $\ket{1}$, which, for simplicity, are taken to be degenerate in energy. A third excited state $\ket{r}$, at an energy of $\omega_r$, mediates a Raman transition between the two ground states. In particular, the transition $\ket{1}\leftrightarrow\ket{r}$ is driven by a highly detuned laser of frequency $\omega_{L_r}$ (detuning $\Delta_r=\omega_{L_r}-\omega_r$), phase $\phi$ and Rabi frequency $\Omega_r$ (taken real), while the other transition $\ket{0}\leftrightarrow\ket{r}$ is driven by the cavity mode. The total spontaneous emission rate from the excited atomic state $\ket{r}$ is $\gamma_r$. The driving laser is detuned from the cavity by an amount $\delta=\omega_{L_r}-\omega$. The squeezed light driving the cavity has a central (or \emph{carrier}) squeezing frequency of $\omega_{L_r}$, i.e., the same as the driving laser. 

\subsection{Master equation}
Setting the zero of energy to be the energy of the (degenerate) ground states $\ket{1}$ and $\ket{0}$, we may write the master equation for this system, in a non-rotating frame, as
\beq
  \dot\rho=-i\com{H}{\rho}+\lio_{cav}\rho+\lio_{spon}\rho,
\eeq
where 
\beq
  H&=&H_{cav}+H_{atom}+H_{atom/laser}+H_{atom/cav},
\aeq
  H_{cav}&=&\omega a^\dag a,
\aeq
  H_{atom}&=&\omega_r\proj{r},
\aeq
  H_{atom/laser}&=&\frac{\Omega_r}{2}\brar{
  e^{-i\phi}e^{-i\omega_{L_r}t}\op{r}{1}
  +e^{i\phi}e^{i\omega_{L_r}t}\op{1}{r}},
\aeq
  H_{atom/cav}&=&g\op{r}{0}a+g\op{0}{r}\ad,
\eeq
and
\beq
  \label{eq:a3:Lcav}
  \lio_{cav}\rho&=&
  \kappa(1+N)\damping{a}{\rho}{a^\dag}
\nel
  &&+\kappa N\damping{a^\dag}{\rho}{a}
\nel
  &&+\kappa Me^{-2i\omega_{L_r}t}\damping{\ad}{\rho}{\ad}
\nel
  &&+\kappa\conj{M}e^{2i\omega_{L_r} t}\damping{a}{\rho}{a}.
\eeq
Note that throughout this paper we use natural units such that $\hbar=1$. Here we assume that the squeezing bandwidth dominates the decay rate $\kappa$ of the cavity, enabling us to write the damping of the cavity in the standard form of~(\ref{eq:a3:Lcav}). For clarity, we will assume in this section that the term $\lio_{spon}\rho$ describing atomic spontaneous emission may be neglected. In section~\ref{sec:a4} the effects of spontaneous emission will be included.  

\subsection{Adiabatic elimination of excited states}
\label{sec:a3:adelim}
To investigate the essential dynamics of this system, we assume that the detuning of the light field from the excited atomic state is very large, i.e.,
\beq
  \modulus{\Delta_r}\gg\kappa,\modulus{\Omega_r},\modulus{g},\gamma_r,
\eeq
so that the state $\ket{r}$ may be adiabatically eliminated. This results in a master equation (with the cavity in a frame rotating at the laser frequency) of the form
\beq
  \dot\rho=-i\com{H}{\rho}+\lio_{cav}\rho,
\eeq
where
\beq
  H&=&H_{atom}+H_{cav}+H_{atom/cav},
\aeq
  H_{atom}&=&\frac{\Omega_r^2}{4\Delta_r}\sigpm,
\aeq
  H_{cav}&=&-\delta\ad a,
\aeq
  \label{eq:a3:Hatomcav}
  H_{atom/cav}&=&
  \frac{g\Omega_r}{2\Delta_r}\brar{e^{i\phi}\sigp a+e^{-i\phi}\sigm\ad}
  +\frac{g^2}{\Delta_r}\sigmp\ad a,
\eeq
and
\beq
  \label{eq:a3:Lcav2}
  \lio_{cav}\rho&=&
  \kappa(1+N)\damping{a}{\rho}{a^\dag}
\nel
  &&+\kappa N\damping{a^\dag}{\rho}{a}
\nel
  &&+\kappa M\damping{\ad}{\rho}{\ad}
\nel
  &&+\kappa\conj{M}\damping{a}{\rho}{a}.
\eeq
Note that the ``raising'' and ``lowering'' operators between the atomic ground states are defined as $\sigp=\op{1}{0}$ and $\sigm=\op{0}{1}$.

With the excited state eliminated we have an effective two-level system whose interaction with the cavity mode is characterized by the parameters
\beq
  \beta_r&=&\frac{g\Omega_r}{2\Delta_r},
\aeq
  \eta_r&=&\frac{g^2}{\Delta_r}.
\eeq
Here $\beta_r$ is the coupling constant of the effective two-level system to the cavity mode and $\eta_r$ is the ac-Stark shift induced in ground state $\ket{0}$ per cavity photon.  

\subsection{Adiabatic elimination of cavity}
Examining~(\ref{eq:a3:Hatomcav}) we now see that the interaction between the cavity mode and the atom is nearly in the Jaynes-Cummings form, except for the presence of an extra term $\eta_r\sigmp\ad a$. To understand the effect of this term on the atomic ground states we adiabatically eliminate the cavity mode by assuming that its decay rate $\kappa$ is larger than its coupling to the effective two-level atomic system. The time-scale assumptions we have made are now
\beq
  \modulus{\Delta_r}&\gg&\kappa,\modulus{\Omega_r},\modulus{g},\gamma_r,
\aeq
  \kappa&\gg&\modulus{\beta_r},\modulus{\eta_r}.
\eeq

To simplify the adiabatic elimination, we ensure that all of the cavity operators coupling to the atom have zero mean by defining a new operator
\beq
  \breve{n}&=&\ad a-N.
\eeq
In order to eliminate systematic motion of the cavity mode at frequency $\delta$, and also of the atomic ground states due to their effective level shifts, we make a further transformation to a new interaction picture relative to the Hamiltonian term:
\beq
  H_0=-\delta\ad a+\frac{\Omega_r^2}{4\Delta_r}\sigpm+\eta_rN\sigmp.
\eeq
After these minor transformations, the new master equation becomes
\beq
  \dot\rho=-i\com{H_{atom/cav}}{\rho}+\lio_{cav}\rho,
\eeq
with
\beq
  H_{atom/cav}=
  \beta_r\brar{e^{i\phi}e^{i\delta t}e^{i\alpha t}\sigp a
  +e^{-i\phi}e^{-i\delta t}e^{-i\alpha t}\sigm\ad}
  +\eta_r\sigmp\breve{n},
\eeq
and
\beq
  \lio_{cav}\rho&=&
  \kappa(1+N)\damping{a}{\rho}{a^\dag}
\nel
  &&+\kappa N\damping{a^\dag}{\rho}{a}
\nel
  &&+\kappa Me^{-2i\delta t}\damping{\ad}{\rho}{\ad}
\nel
  &&+\kappa\conj{M}e^{2i\delta t}\damping{a}{\rho}{a}.
\eeq
We have introduced a new parameter
\beq
  \label{eq:a3:alpha}
  \alpha=\frac{\Omega_r^2}{4\Delta_r}-\frac{g^2N}{\Delta_r}
  =\frac{\beta_r^2}{\eta_r}-\eta_rN,
\eeq
which describes the difference in the level shifts experienced by the ground states~$\ket{0}$ and~$\ket{1}$.

We may now perform the adiabatic elimination of the cavity (following the method used, e.g., in~\cite{Par99/2}) to obtain a master equation for the atomic ground states alone:
\beq
  \label{eq:a3:me}
  \dot{\rho}&=&
  +\beta_r^2\brar{\frac{N+1}{\kappa-i\alpha-i\delta}}
  \brac{\sigm\rho\sigp-\sigpm\rho}
\nel
  &&+\beta_r^2\brar{\frac{N+1}{\kappa+i\alpha+i\delta}}
  \brac{\sigm\rho\sigp-\rho\sigpm}
\nel
  &&+\beta_r^2\brar{\frac{N}{\kappa+i\alpha+i\delta}}
  \brac{\sigp\rho\sigm-\sigmp\rho}
\nel
  &&+\beta_r^2\brar{\frac{N}{\kappa-i\alpha-i\delta}}
  \brac{\sigp\rho\sigm-\rho\sigmp}
\nel
  &&+\beta_r^2\brar{\frac{-\kappa M}{\kappa-i\delta}}
  \frac{e^{2i\phi}e^{2i\alpha t}}{\kappa+i\alpha-i\delta}
  \brac{2\sigp\rho\sigp}
\nel
  &&+\beta_r^2\brar{\frac{-\kappa\conj{M}}{\kappa+i\delta}}
  \frac{e^{-2i\phi}e^{-2i\alpha t}}{\kappa-i\alpha+i\delta}
  \brac{2\sigm\rho\sigm}
\nel
  &&+\eta_r^2\brar{\frac{1}{2\kappa}}
  \brar{N(N+1)+\frac{\kappa^2M\conj{M}}{\kappa^2+\delta^2}}
\nel
  &&\qquad\brac{2\sigmp\rho\sigmp-\sigmp\rho-\rho\sigmp}.
\eeq

\subsection{Comparison to two-level atom}
We are now in a position to analyze the differences between a two-level atom in a squeezed vacuum and the three-level $\Lambda$-system interacting with squeezed light driving the cavity. 

First of all we may compare the master equations~(\ref{eq:a3:me})
and~(\ref{eq:a2:me}). There is a time dependence in master equation~(\ref{eq:a3:me}), which can only be removed when parameters are chosen such that $\alpha=0$. This is a prerequisite for keeping the phase of the atomic quadratures constant with respect to the phase of the squeezed light. We see that the detuning~$\delta$ between the cavity and the laser has not provided a degree of freedom to cancel with~$\alpha$ as we might have hoped.

Even if we choose parameters such that~$\alpha$ and~$\delta$ are zero we still do not have a master equation of the appropriate form~(\ref{eq:a2:me}) due to the term that is proportional to~$\eta_r^2$. This term may be interpreted as phase damping of the qubit caused by a coherent scattering process between the atom and the intra-cavity photons. 

To understand further the effect of the term containing $\eta_r^2$ in~(\ref{eq:a3:me}) it is informative to compare the Bloch equations for the $\Lambda$-system with those for a two-level atom in a broadband squeezed vacuum. Taking $\alpha=0$, and choosing for clarity $\delta=0$ and $M$ real, the reduced master equation for the three-level $\Lambda$-system simplifies to
\beq
  \label{eq:a3:simpleme}
  \dot{\rho}&=&
  \frac{\beta_r^2}{\kappa}(N+1)
  \brac{2\sigm\rho\sigp-\sigpm\rho-\rho\sigpm}
\nel
  &&+\frac{\beta_r^2}{\kappa}N
  \brac{2\sigp\rho\sigm-\sigmp\rho-\rho\sigmp}
\nel
  &&-\frac{\beta_r^2}{\kappa}M
  \brac{2\sigp\rho\sigp+2\sigm\rho\sigm}
\nel
  &&+\frac{\beta_r^2}{\kappa}P\brac{2\sigmp\rho\sigmp-\sigmp\rho-\rho\sigmp},
\eeq
where we have defined the constant
\beq
  P=\frac{N(N+1)+M^2}{2N}.
\eeq

The Bloch equations derived from~(\ref{eq:a3:simpleme}) are then
\beq
  \label{eq:a3:bex}
  \ddt\ex{\sigx}&=&-\frac{\beta_r^2}{\kappa}(2N+1+2M+P)\ex{\sigx},
\aeq
  \label{eq:a3:bey}
  \ddt\ex{\sigy}&=&-\frac{\beta_r^2}{\kappa}(2N+1-2M+P)\ex{\sigy},
\aeq
  \label{eq:a3:bez}
  \ddt\ex{\sigz}&=&-2\frac{\beta_r^2}{\kappa}
  -2\frac{\beta_r^2}{\kappa}(2N+1)\ex{\sigz}.
\eeq
These equations may be compared to the Bloch equations~(\ref{eq:a2:bex}--\ref{eq:a2:bez}) for a single two-level atom in squeezed vacuum. We see that the two sets of Bloch equations are of the same form (up to an overall rate) except for the extra term~$P$ present in the phase decay terms for the $\Lambda$-atom in a cavity. This extra factor causes enhanced phase decay in both quadratures but does not affect the populations or their decay rates. 

Full equivalence between the Bloch equations~(\ref{eq:a2:bex}--\ref{eq:a2:bez}) and~(\ref{eq:a3:bex}--\ref{eq:a3:bez}) would be realised if we could choose parameters such that $P\ll2N+1-2M$. However, investigation of this inequality reveals that, at best,~$P$ can be reduced to approximately $20\percent$ of the value of $2N+1-2M$, and then only for very small values of~$M$. Hence, there is always significant phase decay in each quadrature and we conclude that it is not possible to choose a parameter regime which permits the $\Lambda$-atom in a cavity to have similar phase decay dynamics to the equivalent two-level atom. In addition, it is easy to establish that it is not possible to choose values of~$N$ and~$M$ such that $2N+1-2M+P<1$. Thus, with the three-level $\Lambda$-system, it is not possible to reduce phase damping to a degree which permits a quadrature decay rate lower than that obtained when the cavity is damped by ordinary vacuum.

\section{Four-level $\Lambda$-system}
\label{sec:a4}
\newcommand{\Delsq}{\Delta_r^2}
The time dependence in~(\ref{eq:a3:me}) arises from the differing level shifts experienced by states~$\ket{0}$ and~$\ket{1}$ in our effective two level system. This suggests that the addition of another level shift into one of these ground states could provide a degree of freedom with which to cancel the time dependence. A technique to achieve this is to virtually excite a fourth atomic level by driving it selectively from one of the ground states via a second laser field.

\begin{figure}
  \begin{center}
  \includegraphics{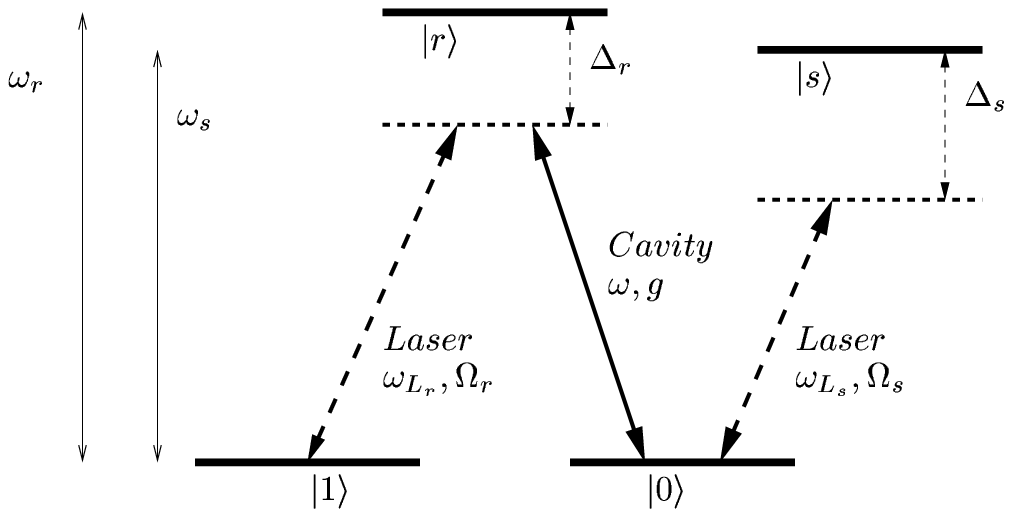}
  \end{center}
  \caption{The atomic levels and transitions for a four-level atom, showing the resonant Raman transition and the virtual excitation of state~$\ket{s}$. The two driving lasers have frequencies $\omega_{L_r}$ and $\omega_{L_s}$, and Rabi frequencies $\Omega_r$ and $\Omega_s$, respectively (both taken real). The excited levels have frequencies $\omega_r$ and $\omega_s$ and spontaneous emission rates~$\gamma_r$ and~$\gamma_s$. The lasers are strongly detuned by $\Delta_r=\omega_{L_r}-\omega_r$ and $\Delta_s=\omega_{L_s}-\omega_s$. As drawn, the detunings $\Delta_r$ and $\Delta_s$ are negative.}
  \label{fig:a4}
\end{figure}

A new atomic level scheme to achieve this is shown in figure~\ref{fig:a4}. 
This configuration has the same basic structure as that shown in figure~\ref{fig:a3}, except for the addition of a new excited state~$\ket{s}$ and a second laser field coupling it to level~$\ket{0}$. We saw in section~\ref{sec:a3} that driving the system off Raman resonance ($\delta\ne0$) was not an effective approach for cancelling the time dependence introduced by the level shifts. Therefore, we now take $\delta=0$ (i.e. drive the transition $\ket{0}\leftrightarrow\ket{1}$ on Raman resonance). The state~$\ket{s}$ is virtually excited from~$\ket{0}$ by a second strongly detuned laser with Rabi frequency~$\Omega_s$ and detuning~$\Delta_s$. It has a spontaneous emission rate~$\gamma_s$. We expect that the laser field~$\Omega_s$ driving the transition~$\ket{0}\leftrightarrow\ket{s}$ will add an additional ac-Stark shift to the ground state~$\ket{0}$. 

We saw for the three-level $\Lambda$-atom that the phase~$\phi$ of the laser~$\Omega_r$ was only of relevance when compared to the phase of the squeezed light, as seen in~(\ref{eq:a3:me}). Thus, for clarity, this phase factor could have been included in~$M$ from the outset and otherwise not considered. We take this approach in our analysis of the four-level $\Lambda$-system. Also the phase of the laser~$\Omega_s$ is of no relevance and has been neglected.
 
For the four-level $\Lambda$-atom we analyze the effects of atomic spontaneous emission (neglected in section~\ref{sec:a3}). We take the spontaneous emission term from excited state~$\ket{r}$ to be described by \commentout{(W102B)}
\beq
  \label{eq:a4:Lspon}
  \lio_{spon}\rho&=&
  \frac{\gamma_r}{2}b_0^2
  \brac{2\op{0}{r}\rho\op{r}{0}-\proj{r}\rho-\rho\proj{r}}
\nel
  &+&\frac{\gamma_r}{2}b_1^2
  \brac{2\op{1}{r}\rho\op{r}{1}-\proj{r}\rho-\rho\proj{r}}.
\eeq
Spontaneous emission from the excited level~$\ket{r}$ can proceed via two channels $\ket{r}\rightarrow\ket{0}$ and $\ket{r}\rightarrow\ket{1}$ with branching ratios~$b_0$ and~$b_1$ respectively, such that $b_0^2+b_1^2=1$. We shall see that we are free to increase $\Delta_s$ so long as the ratio $\frac{\Omega_s}{\Delta_s}$ is kept the same. This means that spontaneous emission from the excited level~$\ket{s}$ can be made arbitrarily small and may safely be neglected. 

\subsection{Adiabatic elimination of excited states}
\label{sec:a4:adelim}
Following the same approach as for the three-level $\Lambda$-system we may again adiabatically eliminate the excited states of the atoms with the timescale conditions
\beq
  \label{eq:a4:adelcond1}
  \modulus{\Delta_r},\modulus{\Delta_s}&\gg&\kappa,
  \modulus{\Omega_r},\modulus{\Omega_s},\modulus{g},\gamma_r,\gamma_s,
\eeq
leading to the master equation for the cavity and the atomic ground states (in a frame rotating at the cavity frequency) of \commentout{(W102F)}
\beq
  \dot\rho=-i\brar{H_{eff}\rho-\adj{H_{eff}}}+\lio_{spon}\rho+\lio_{cav}\rho,
\eeq
where
\beq
  H_{eff}&=&\brar{\frac{\Delta_r-i\frac{\gamma_r}{2}}{\Delsq}}
  \brac{g^2\ad a\sigm\sigp+\frac{\Omega_r^2}{4}\sigp\sigm}
\nel
  &&+\brar{\frac{\Delta_r-i\frac{\gamma_r}{2}}{\Delsq}}\frac{g\Omega_r}{2}
  \brar{a\sigp+\ad\sigm}
  +\frac{\Omega_s^2}{4\Delta_s}\sigm\sigp.
\eeq
The form of $L_{cav}\rho$ is here identical to~(\ref{eq:a3:Lcav2}).

After the adiabatic elimination of the excited states the effect of spontaneous emission from the excited state~$\ket{r}$ is now described by 
\beq
  \label{eq:a4:Lspon2}
  \lio_{spon}\rho&=&\frac{\gamma_r}{\Delsq}
  \brar{D_0\rho\adj{D_0}+D_1\rho\adj{D_1}},
\aeq
  D_0&=&b_0\brar{ga\sigm\sigp+\frac{\Omega_r}{2}\sigm},
\aeq
  D_1&=&b_1\brar{ga\sigp+\frac{\Omega_r}{2}\sigp\sigm},
\eeq
and is comprised of three varieties of spontaneous emission, characterized by the rates
\beq
  \gamma_r\frac{g^2N}{\Delsq},
  \ \gamma_r\frac{g\sqrt{N}\Omega_r}{2\Delsq},
  \ \gamma_r\frac{\Omega_r^2}{4\Delsq}.
\eeq
To simplify the analysis we make the assumption that
\beq
  \label{eq:a4:g2Nsmall}
  \modulus{g}\sqrt{N}\ll\modulus{\Omega_r},
\eeq
and keep only the terms in~(\ref{eq:a4:Lspon2}) that depend on $\Omega_r^2$. This leads to the master equation \commentout{(W102J)}
\beq
  \dot\rho=-i\com{H}{\rho}+L_a\rho+L_{cav}\rho,
\eeq
where
\beq
  H&=&H_a+H_{ac},
\aeq
  \label{eq:a4:Ha}
  H_a&=&\frac{g^2}{\Delta_r}N\sigmp+\frac{\Omega_r^2}{4\Delta_r}\sigpm
  +\frac{\Omega_s^2}{4\Delta_s}\sigmp,
\aeq
  H_{ac}&=&
  \frac{g^2}{\Delta_r}\breve{n}\sigmp
  +\frac{g\Omega_r}{2\Delta_r}\brar{a\sigp+\ad\sigm},
\eeq
and
\beq
  \lio_a\rho&=&
  b_0^2\frac{\gamma_r}{2}\frac{\Omega_r^2}{4\Delta_r^2}
  \damping{\sigm}{\rho}{\sigp}
\nel
  &&+b_1^2\frac{\gamma_r}{2}\frac{\Omega_r^2}{4\Delta_r^2}
  \brar{2\sigpm\rho\sigpm-\sigpm\rho-\rho\sigpm}.
\eeq
Note that the action of the second driving laser has been to add an additional level shift into the ground state~$\ket{0}$, given by the third term in~(\ref{eq:a4:Ha}).

\subsection{Adiabatic elimination of cavity}
The further adiabatic elimination of the cavity requires parameters satisfying the bad-cavity limit
\beq
  \label{eq:a4:adelcond2}
  \kappa&\gg&\modulus{\beta_r},\modulus{\eta_r}.
\eeq
Again we ensure that $\alpha=0$ in order to balance the level shifts in the effective ground states and to ensure significant atomic squeezing. The definition of~$\alpha$ is modified somewhat from~(\ref{eq:a3:alpha}) and becomes \commentout{(W102K)}
\beq
  \label{eq:a4:sealpha}
  \alpha&=&\frac{\Omega_r^2}{4\Delta_r}
  -\frac{\Omega_s^2}{4\Delta_s}
  -\frac{g^2}{\Delta_r}N.
\eeq

With the condition $\alpha=0$, the adiabatic elimination of the cavity leads to a master equation for the atomic system alone \commentout{(W106O)}
\beq
  \ddt{\rho}&=&\frac{\beta_r^2}{\kappa}
  \brar{N+1+\frac{b_0^2}{2C}}\brac{\dampingnobr{\sigm}{\rho}{\sigp}}
\nel
  &&+\frac{\beta_r^2}{\kappa}N
  \brac{\dampingnobr{\sigp}{\rho}{\sigm}}
\nel
  &&-\frac{\beta_r^2}{\kappa}M\brac{2\sigp\rho\sigp}
  -\frac{\beta_r^2}{\kappa}\conj{M}\brac{2\sigm\rho\sigm}
\nel
  &&+\frac{\beta_r^2}{\kappa}P\brac{2\sigpm\rho\sigpm-\sigpm\rho-\rho\sigpm},
\eeq
where we have defined the constants
\beq
  \label{eq:a4:C}
  C&=&\frac{g^2}{\kappa\gamma_r},
\aeq
  \label{eq:a4:P}
  P&=&\frac{2g^2}{\Omega_r^2}\brac{N(N+1)+M^2}
  +\frac{b_1^2}{2C}.
\eeq
We may make some observations from this master equation. First of all, the overall rate of the evolution of the density matrix is set by the value of $\beta_r^2/\kappa$. Therefore increasing the detuning $\Delta_r$ merely serves to slow the evolution to the steady state, but does not otherwise change the overall dynamics. Also, the last line of the master equation represents phase damping due both to spontaneous emission (from~$\ket{r}$ to~$\ket{1}$) and to elastic photon scattering into the cavity mode. This scattering is naturally dependent both on the degree of squeezing~$M$ as well as the mean photon number~$N$ as seen in~(\ref{eq:a4:P}).

\subsection{Comparison to two-level atom}
The Bloch equations for this master equation, taking $M$ to be real, are \commentout{(W108A,B,AP,AQ,AR)}
\beq
  \label{eq:a4:bex}
  \ddt{\ex{\sigx}}&=&-\Gamma_x\ex{\sigx},
\aeq
  \label{eq:a4:bey}
  \ddt{\ex{\sigy}}&=&-\Gamma_y\ex{\sigy},
\aeq
  \label{eq:a4:bez}
  \ddt{\ex{\sigz}}&=&-\Gamma-\Gamma_z\ex{\sigz},
\eeq
with damping constants
\beq
  \Gamma_x&=&\frac{\beta_r^2}{\kappa}\brar{2N+1+2M+D},
\aeq
  \Gamma_y&=&\frac{\beta_r^2}{\kappa}\brar{2N+1-2M+D},
\aeq
  \Gamma_z&=&2\frac{\beta_r^2}{\kappa}\brac{2N+1+\frac{b_0^2}{2C}},
\aeq
  \label{eq:a4:Gamma}
  \Gamma&=&2\frac{\beta_r^2}{\kappa}\brac{1+\frac{b_0^2}{2C}},
\eeq
and where we have additionally defined
\beq
  D=\frac{2g^2}{\Omega_r^2}\brac{N(N+1)+M^2}
  +\frac{1}{2C}.
\eeq
The conditions under which these Bloch equations have a similar form (up to an overall rate) as equations (\ref{eq:a2:bex}--\ref{eq:a2:bez}) are that a)~$D$ is much less than $2N+1-2M$, and b) the first term dominates the second term in~(\ref{eq:a4:Gamma}). These conditions may be written as \commentout{(W108AS,AT,AU)}
\beq
  \label{eq:a4:nophasedamping}
  \frac{g^2}{\Omega_r^2}&\ll&\frac{N+\half-M}{N(N+1)+M^2},
\aeq
  \label{eq:a4:se:cond2}
  C&\gg&\frac{1}{2\brar{2N+1-2M}},
\aeq
  \label{eq:a4:se:cond3}
  C&\gg&\frac{b_0^2}{2}.
\eeq
Note that~(\ref{eq:a4:se:cond2}) and~(\ref{eq:a4:se:cond3}) amount to the condition of strong coupling in cavity QED, made somewhat more stringent due to the inhibited atomic decay rate associated with the squeezed reservoir interaction (i.e., due to the factor $2N+1-2M<1$).

The addition of the fourth level has led to a significant improvement. It is now possible to satisfy the condition $\alpha=0$, while simultaneously satisfying conditions (\ref{eq:a4:adelcond1}, \ref{eq:a4:adelcond2}, \ref{eq:a4:nophasedamping}, \ref{eq:a4:se:cond2}, \ref{eq:a4:se:cond3}). In particular, this may be achieved with the following steps: 
\begin{enumerate}
\item Choose basic cavity QED parameters $g$, $\kappa$ and $\gamma$ such that the strong coupling conditions~(\ref{eq:a4:se:cond2}) and~(\ref{eq:a4:se:cond3}) are satisfied for the desired values of $N$ and $M$. 
\item Choose a value of~$\Omega_r$ sufficiently large that the phase damping is negligible and that spontaneous emission arises predominantly from the laser excitation of the atom rather than from cavity excitation (i.e. choose~$\Omega_r$ to satisfy both equations~(\ref{eq:a4:nophasedamping}) and~(\ref{eq:a4:g2Nsmall})).
\item Choose~$\Delta_r$ so that our condition~(\ref{eq:a4:adelcond2}) for the adiabatic elimination of the cavity is satisfied.
\item Finally, choose~$\Delta_s$ and~$\Omega_s$ so that $\alpha=0$ (as is required to balance the level shifts in the ground states). Note that spontaneous emission from the excited state~$\ket{s}$ occurs at a rate of $\gamma_s\frac{\Omega_s^2}{4\Delta_s^2}$, whereas $\alpha$ depends on the term $\frac{\Omega_s^2}{4\Delta_s}$, so spontaneous emission from~$\ket{s}$ can always be neglected with a sufficiently large choice of~$\Delta_s$.
\end{enumerate}
With these choices of parameters we can make the behaviour of the ground states of a 4-level $\Lambda$-system in a cavity driven by squeezed light identical to the behaviour of a free two-level atom interacting exclusively with squeezed light. This is the main result of this paper.
 
\subsection{Probe transmission spectrum}
An identifiable signature of inhibited atomic phase decay is a feature of subnatural linewidth in the spectrum of a weak coherent probe field transmitted through the cavity~\cite{Par93,Dal99,Tur98/2}. When using a single probe field, the probe transmission spectrum is a function of both quadrature decay rates~$\Gamma_x$ and~$\Gamma_y$. This can to some extent diminish the contribution from the particular quadrature that has reduced noise. 

Therefore, in this section we introduce a technique employing a pair of probe fields which are symmetrically distributed about the cavity frequency. By using two probe fields we can cancel the contribution from the noisy quadrature and measure a probe transmission spectrum which depends only upon the quadrature with reduced fluctuations (e.g.~$\Gamma_y$ when~$M$ is real and positive). We consider the mean field amplitude transmitted through the cavity when driven by a pair of weak probe fields, i.e., \commentout{(W109)}
\beq
  H_{probe}&=&i\brac{\mathcal{E}(t)\ad-\conj{\mathcal{E}}(t)a},
\aeq
  \mathcal{E}(t)&=&\mathcal{E}_+e^{-i\brar{\omega+\nu}t}
  +\mathcal{E}_-e^{-i\brar{\omega-\nu}t}.
\eeq
Here~$\omega$ is the cavity frequency, and~$\mathcal{E}_+$ and~$\mathcal{E}_-$ are the weak complex driving amplitudes of the two probe fields which are offset by frequencies~$\nu$ and~$-\nu$ respectively from~$\omega$. Of course, the normal transmission spectrum for a single probe field can be recovered from the more general case that we consider by taking $\mathcal{E}_-=0$.

We assume here that $\alpha=0$ and that equations (\ref{eq:a4:adelcond1}, \ref{eq:a4:adelcond2}, \ref{eq:a4:nophasedamping}, \ref{eq:a4:se:cond2}, \ref{eq:a4:se:cond3}) are satisfied. Then, the field amplitude is given by (see Appendix for details of derivation) \commentout{(W115G)}
\beq
  \label{eq:a4:exa}
  \ex{a(t)}&=&
  e^{-i\nu t}A_p^+(\nu)+e^{i\nu t}A_p^-(\nu),
\aeq
  \label{eq:a4:App}
  A_p^+(\nu)&=&\frac{\mathcal{E}_+}{\kappa-i\nu}
  +\frac{\beta_r^2}{2\kappa^2}\ex{\sigz}
  \brar{\frac{\mathcal{E}_++\conj{\mathcal{E}_-}}{\Gamma_x-i\nu}
  +\frac{\mathcal{E}_+-\conj{\mathcal{E}_-}}{\Gamma_y-i\nu}},
\aeq
  \label{eq:a4:Apm}
  A_p^-(\nu)&=&\frac{\mathcal{E}_-}{\kappa+i\nu}
  +\frac{\beta_r^2}{2\kappa^2}\ex{\sigz}
  \brar{\frac{\mathcal{E}_-+\conj{\mathcal{E}_+}}{\Gamma_x+i\nu}
  +\frac{\mathcal{E}_--\conj{\mathcal{E}_+}}{\Gamma_y+i\nu}}.
\eeq
Here we have defined probe transmission amplitudes for the upper and lower sidebands as $A_p^+(\nu)$ and $A_p^-(\nu)$, respectively.

In practice, these amplitudes can be measured precisely using heterodyne detection~\cite{Tur98/2}. We can see from equations~(\ref{eq:a4:App}) and~(\ref{eq:a4:Apm}) that variation of the relative phase of~$\mathcal{E}_+$ and~$\mathcal{E}_-$, while keeping $\modulus{\mathcal{E}_+}=\modulus{\mathcal{E}_-}$, results in the transmission spectrum for both the upper and lower sidebands being selectively sensitive to only one of the quadrature decay rates~$\Gamma_x$ or~$\Gamma_y$. In particular, for $\mathcal{E}_+=\mathcal{E}_-=\mathcal{E}$, we have
\beq
  A_p^\pm(\nu)&=&\frac{\mathcal{E}}{\kappa\mp i\nu}
  +\frac{\beta_r^2}{\kappa^2}\ex{\sigz}
  \frac{\mathcal{E}}{\Gamma_x\mp i\nu},
\eeq
while for $\mathcal{E}_+=-\mathcal{E}_-=\mathcal{E}$,
\beq
  A_p^\pm(\nu)&=&\frac{\pm\mathcal{E}}{\kappa\mp i\nu}
  \pm\frac{\beta_r^2}{\kappa^2}\ex{\sigz}
  \frac{\mathcal{E}}{\Gamma_y\mp i\nu}.
\eeq

\begin{figure}
  \begin{center}
  \includegraphics[scale=0.6]{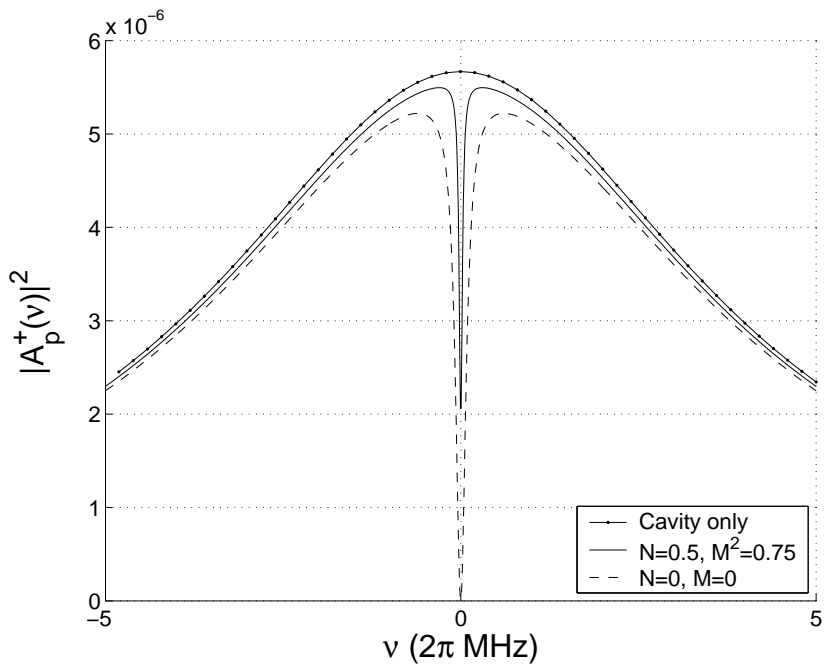}
  \end{center}
  \caption{Upper sideband probe transmission spectrum $\modsquared{A_p^+(\nu)}$ for a 4-level $\Lambda$-atom in a cavity driven by squeezed light. Both the upper and lower sidebands are probed, with $\mathcal{E}_+/(2\pi)=0.01$ MHz and $\mathcal{E}_-/(2\pi)=-0.01$ MHz. The topmost curve shows the response expected for a cavity containing no atoms. The lower two curves show the upper sideband transmission spectrum for ideal squeezed driving ($M=\sqrt{N(N+1}$) and ordinary vacuum ($N=M=0$), with $g/(2\pi)=24$ MHz, $\kappa/(2\pi)=4.2$ MHz and $\gamma_r/(2\pi)=5.2$ MHz as realised in a recent single-atom trapping experiment~\cite{McK02P}. Other relevant parameters are $g/\Omega_r=0.1$, $\Omega_r/\Delta_r=0.05$ and $\alpha=0$, which give $\beta_r/(2\pi)=0.6$ MHz and $\eta_r/(2\pi)=0.12$ MHz. Note that we have taken branching ratios for atomic spontaneous emission to be $b_0=b_1=1/\sqrt{2}$.} 
  \label{fig:usb}
\end{figure}
\begin{figure}
  \begin{center}
  \includegraphics[scale=0.6]{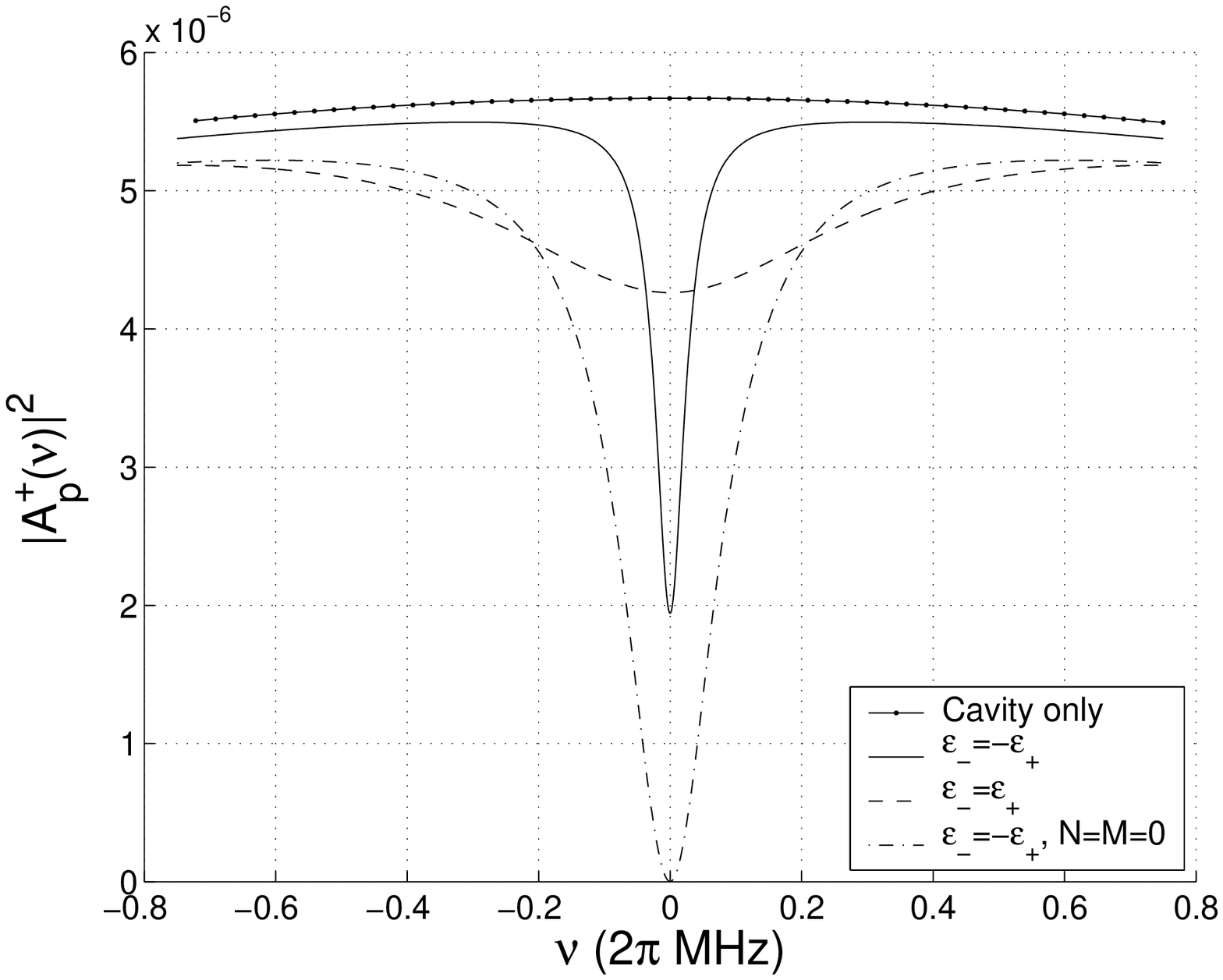}
  \end{center}
  \caption{Magnified version of figure~\ref{fig:usb}. Upper sideband probe transmission spectrum $\modsquared{A_p^+(\nu)}$ for a 4-level $\Lambda$-atom in a cavity driven by squeezed light. This graph has a smaller range of frequencies than figure~\ref{fig:usb} and shows curves for different relative phases of $\mathcal{E}_+$ and $\mathcal{E}_-$, with $\mathcal{E}_+/(2\pi)=0.01$ MHz. Variation in the phase of $\mathcal{E}_+$ and $\mathcal{E}_-$ results in the difference in linewidths seen between the solid and dashed lines. The lowest curve shows the natural linewidth associated with the cavity driven by ordinary vacuum.} 
  \label{fig:usb2}
\end{figure}
We demonstrate this effect by plotting the upper sideband probe transmission spectrum in figures~\ref{fig:usb} and~\ref{fig:usb2}. To generate these figures we have used realistic parameters for $g$, $\kappa$ and $\gamma_r$ obtained in a recent cavity-QED single-atom trapping experiment~\cite{McK02P}. Our use of a $\Lambda$-system means that the characteristic rate $\beta_r^2/\kappa$ for the evolution of the system is substantially smaller than~$\kappa$. This means that the dip in the transmission spectrum around~$\nu=0$ is a narrow feature compared to the broad transmission spectrum that would be seen for the cavity alone (topmost line in figures~\ref{fig:usb} and~\ref{fig:usb2}). With a modest amount of ideal squeezing ($N=0.5$, $M=\sqrt{0.75}$) we see a significant narrowing of the central feature in the probe spectrum (for $\mathcal{E}_+=-\mathcal{E}_-$) as a result of the quadrature noise reduction.

\begin{figure}
  \begin{center}
  \includegraphics[scale=0.6]{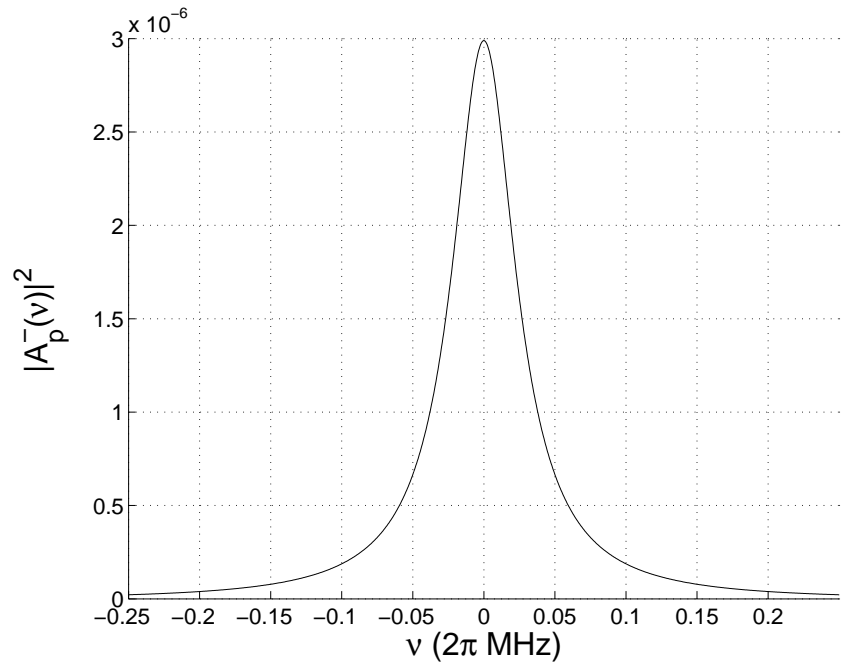}
  \end{center}
  \caption{Lower sideband probe transmission spectrum $\modsquared{A_p^-(\nu)}$ for a 4-level $\Lambda$-atom in a cavity driven by squeezed light characterized by $N=0.5$ and $M^2=0.75$. Only the upper probe sideband is driven with $\mathcal{E}_+/(2\pi)=0.01$ MHz and $\mathcal{E}_-=0$. Other parameters are the same as in figure~\ref{fig:usb}.}
  \label{fig:lsb}
\end{figure}

It may also be seen from equations (\ref{eq:a4:exa}--\ref{eq:a4:Apm}) that when the incident probe field contains only the upper sideband (i.e., when $\mathcal{E}_+\ne0$, $\mathcal{E}_-=0$) then, provided $M\ne0$, a finite amplitude lower sideband develops in the transmission. In particular, if $\mathcal{E}_-=0$, and $\mathcal{E}_+=\conj{\mathcal{E}_+}=\mathcal{E}$, then
\beq
  A_p^-(\nu)&=&\frac{\beta_r^2}{2\kappa^2}\ex{\sigz}\mathcal{E}
  \brar{\frac{1}{\Gamma_x+i\nu}-\frac{1}{\Gamma_y+i\nu}}
\nel
  &=&\mathcal{E}\frac{\beta_r^2}{2\kappa^2}\ex{\sigz}
  \brar{\frac{\Gamma_y-\Gamma_x}{\brar{\Gamma_x+i\nu}\brar{\Gamma_y+i\nu}}}
\nel
  &=&-2\mathcal{E}\frac{\beta_r^4}{\kappa^3}\ex{\sigz}
  \frac{M}{\brar{\Gamma_x+i\nu}\brar{\Gamma_y+i\nu}},
\eeq
since
\beq
  \Gamma_y-\Gamma_x=-\frac{4\beta_r^2M}{\kappa}.
\eeq
The dependence on~$M$ indicates that this effect is a consequence of correlations in the input squeezed light. Examples of the lower sideband transmission spectrum for this case are shown in figure~\ref{fig:lsb}. Importantly, the width of the spectral feature is again subnatural (compared to the ``ordinary vacuum'' width $\frac{2\beta_r^2}{\kappa}/(2\pi)=0.17$ MHz).

\begin{figure}
  \begin{center}
  \includegraphics[scale=0.6]{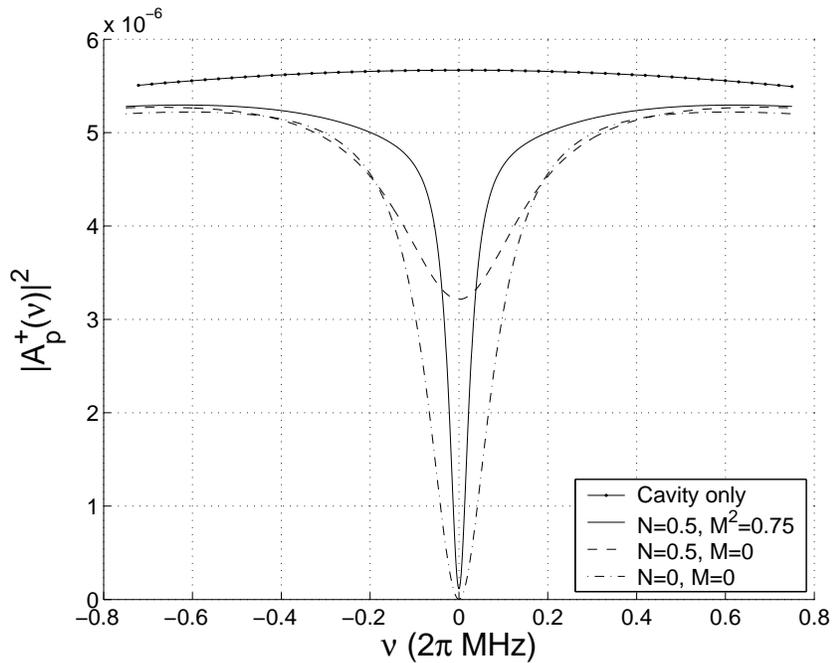}
  \end{center}
  \caption{Upper sideband probe transmission spectrum $\modsquared{A_p^+(\nu)}$ for a 4-level $\Lambda$-atom in a cavity driven by squeezed light. Only the upper probe sideband is driven with $\mathcal{E}_+/(2\pi)=0.01$ MHz and $\mathcal{E}_-=0$. Other parameters are the same as in figure~\ref{fig:usb}. }
  \label{fig:pts}
\end{figure}

Finally, for completeness, in figure~\ref{fig:pts} we plot the standard probe transmission spectrum, i.e., $\modsquared{A_p^+(\nu)}$ for $\mathcal{E}_+\ne0$ and $\mathcal{E}_-=0$. The structure of the central feature is now determined by both~$\Gamma_x$ and~$\Gamma_y$, but again the narrowing due to quadrature noise reduction is clearly visible.

\section{Conclusions}
We have proposed a simple system, based on a 4-level $\Lambda$-atom in a cavity, which has the same interaction with squeezed light driving the cavity as a free 2-level atom interacting exclusively with broadband squeezed light. 

We have identified two major conditions that must be satisfied for this equivalence to be established. Firstly, the level shifts which are induced in the two atomic ground states by the laser driving the Raman transition and by the intra-cavity photons must be balanced. This keeps the quadratures of the atomic ground states in phase with the squeezing incident on the cavity. The fourth atomic level and its associated driving laser are used to achieve this balancing. Secondly, a parameter regime must also be chosen in which phase-damping of the atoms is negligible. The level shift associated with intra-cavity photons causes this phase-damping, which may be considered as a two-photon elastic scattering process between the atomic ground state and the photons in the cavity. With only a 3-level atomic system there is not sufficient flexibility to simultaneously satisfy both these conditions.

An optical cavity is used, in the (``effective'') bad-cavity limit, to create an atom that is effectively one-dimensional. Therefore, we need only squeeze the modes into which the cavity mode decays and do not need to squeeze the full 4$\pi$ of modes with which an atom in free space interacts. The bandwidth of the squeezing must only exceed the bandwidth of the cavity. These are substantially simpler technical challenges.  

The 4-level system considered in this paper might be used to demonstrate the inhibited quadrature decay rates associated with atomic squeezing. Therefore, we have considered a technique to assist in the detection of the line-narrowing which is the direct signature of atomic squeezing. By employing an appropriate pair of probe fields that are symmetrically placed around the resonant frequency of the cavity we may measure a probe transmission spectrum that is dependent only on the reduced atomic quadrature decay rate.

Finally, we note that the basic conditions elucidated in this paper are also applicable to systems containing more than one atom. The requirements to balance the level shifts of the effective ground states and to minimize phase-damping should persist when attempting to create spin-squeezing in larger atomic ensembles. The squeezing created in the atomic ground states is in principle long-lived and will endure when all of the light sources are turned off, making systems of more than one atom particularly suitable for the creation of entangled states of multiple atoms.

\ack
This work was supported in part by the Marsden Fund of the Royal Society of New Zealand.

\appendix
\section*{Appendix. Probe transmission spectrum}
\setcounter{section}{1}
In this appendix we outline the approach taken to calculate the probe transmission spectrum for the 4-level $\Lambda$-atom in a cavity, when driven by squeezed light and by a pair of weak coherent probe fields symmetrically distributed about the cavity frequency. 

The interaction-picture perturbation of our system from its steady-state value~$\rho_{eq}$ is given by~\cite{bookPAA} \commentout{(W115A)}
\beq
  \rho(t)=\rho_{eq}-i\int_{-\infty}^tdt^\prime
  \ \com{H_{probe}\brar{t^\prime}}{\rho_{eq}},
\eeq
where the interaction of the two probe fields with the cavity mode is described by \commentout{(W115C)}
\beq
  H_{probe}(t)&=&i\brac{\mathcal{E}(t)\ad-\conj{\mathcal{E}}(t)a},
\aeq
  \mathcal{E}(t)&=&\mathcal{E}_+e^{-i\nu t}
  +\mathcal{E}_-e^{i\nu t}.
\eeq
From this it may be shown that \commentout{(W115F)}
\beq
  \label{eq:app:exa}
  \ex{a(t)}&=&e^{-i\nu t}\intzi d\tau\ e^{i\nu\tau}A(\tau)
  +e^{i\nu t}\intzi d\tau\ e^{-i\nu\tau}B(\tau),
\aeq
  \label{eq:app:A}
  A(\tau)&=&\mathcal{E}_+\ex{\com{a(\tau)}{\ad}}
  -\mathcal{E}^\ast_-\ex{\com{a(\tau)}{a}},
\aeq
  \label{eq:app:B}
  B(\tau)&=&\mathcal{E}_-\ex{\com{a(\tau)}{\ad}}
  -\mathcal{E}^\ast_+\ex{\com{a(\tau)}{a}}.
\eeq

We must now evaluate the commutators of the two-time correlation functions of the cavity modes. The quantum Langevin equation for the evolution of~$a(t)$ is
\beq
  \label{eq:app:dadt}
  \ddt{a(t)}=-\kappa a(t)+g\sigm(t)+\sqrt{2\kappa}a_{in}(t),
\eeq
where $a_{in}(t)$ is a quantum noise operator satisfying
\beq
  \com{a_{in}(t)}{\ad_{in}(t^\prime)}=\delta\brar{t-t^\prime}.
\eeq
The formal solution to~(\ref{eq:app:dadt}), in the limit of~$\kappa$ large and~$\sigm$ slowly varying, is
\beq
  \label{eq:app:a(t)}
  a(t)=\frac{\beta_r}{\kappa}\sigm(t)+\sqrt{2\kappa}
  \int_0^tdt^\prime\ e^{-\kappa(t-t^\prime)}a_{in}(t^\prime).
\eeq
Evaluation of the commutators in~(\ref{eq:app:A}) and~(\ref{eq:app:B}), using~(\ref{eq:app:a(t)}), gives \commentout{(W116D,W111E)}
\beq
  \ex{\com{a(t)}{a(t^\prime)}}&=&
  \frac{\beta_r^2}{\kappa^2}\ex{\com{\sigm(t)}{\sigm(t^\prime)}}
\nel
  &&+\beta_r\sqrt{\frac{2}{\kappa}}\int_0^td\tau
  \ e^{-\kappa(t-\tau)}
  \ex{\com{a_{in}(\tau)}{\sigm(t^\prime)}}
\nel
  &&+\beta_r\sqrt{\frac{2}{\kappa}}\int_0^{t^\prime}d\tau^\prime
  \ e^{-\kappa(t^\prime-\tau^\prime)}
  \ex{\com{\sigm(t)}{a_{in}(\tau^\prime)}}
\eeq
and
\beq
  \ex{\com{a(t)}{\ad(t^\prime)}}&=&
  \frac{\beta_r^2}{\kappa^2}\ex{\com{\sigm(t)}{\sigp(t^\prime)}}
\nel
  &&+\beta_r\sqrt{\frac{2}{\kappa}}\int_0^td\tau
  \ e^{-\kappa(t-\tau)}
  \ex{\com{a_{in}(\tau)}{\sigp(t^\prime)}}
\nel
  &&+\beta_r\sqrt{\frac{2}{\kappa}}\int_0^{t^\prime}d\tau^\prime
  \ e^{-\kappa(t^\prime-\tau^\prime)}
  \ex{\com{\sigm(t)}{\ad_{in}(\tau^\prime)}}.
\eeq
We may now use the standard input and output relation, which holds for any atomic operator~$x(t)$,
\beq
  \com{x(t)}{a_{in}(t^\prime)}=
  -u(t-t^\prime)\beta_r\sqrt{\frac{2}{\kappa}}
  \com{x(t)}{\sigm(t^\prime)},
\eeq
to express the expectation values of two-time correlations of the cavity operator~$a$ in terms of atomic correlations: \commentout{(W116E,F)}
\beq
  \label{eq:app:aa}
  \ex{\com{a(\tau)}{a}}&=&-\frac{\beta_r^2}{\kappa^2}
  \ex{\com{\sigm(\tau)}{\sigm}},
\aeq
  \label{eq:app:aad}
  \ex{\com{a(\tau)}{\ad}}&=&e^{-\kappa\tau}-\frac{\beta_r^2}{\kappa^2}
  \ex{\com{\sigm(\tau)}{\sigp}}.
\eeq
These in turn may be calculated from (\ref{eq:a4:bex}--\ref{eq:a4:bez}) to be \commentout{(W107R,S)}
\beq
  \label{eq:app:smm}
  \ex{\com{\sigm(\tau)}{\sigm}}&=&\half\ex{\sigz}
  \brar{e^{-\Gamma_x\tau}-e^{-\Gamma_y\tau}},
\aeq
  \label{eq:app:smp}
  \ex{\com{\sigm(\tau)}{\sigp}}&=&-\half\ex{\sigz}
  \brar{e^{-\Gamma_x\tau}+e^{-\Gamma_y\tau}}.
\eeq
Substituting equations~(\ref{eq:app:aa}--\ref{eq:app:smp}) into~(\ref{eq:app:exa}--\ref{eq:app:B}) leads to equations~(\ref{eq:a4:exa}--\ref{eq:a4:Apm}).

\end{document}